\documentclass[epj]{svjour} 
\journalname{}
\usepackage{amsmath}
\usepackage{amssymb}
\usepackage{graphicx}
\usepackage{cite}

\newcommand{\qexpect}[1]{\langle #1 \rangle_{q}}%
\newcommand{\qoneexpect}[1]{\langle #1 \rangle_{q=1}}%
\newcommand{\trace}[1]{\mathrm{Tr}\left( #1\right)} %
\newcommand{\betaph}{\beta_{\mathrm{ph}}}
\newcommand{\betadist}{\beta_{\mathrm{dist}}}
\newcommand{\Tdist}{T_{\mathrm{dist}}} 
\newcommand{\Tph}{T_{\mathrm{ph}}} 

\begin{document}
\title{Momentum distribution and correlation for a free scalar field in the Tsallis nonextensive statistics based on density operator}
\author{Masamichi Ishihara \thanks{\email{m\_isihar@koriyama-kgc.ac.jp}}}
\institute{Department of Human Life Studies, Koriyama Women's University, Koriyama, Fukushima, 963-8503, Japan}

\abstract{
We derived the expression of the normalized $q$-expectation value based on the density operator to the order $1-q$ 
with the physical temperature in the Tsallis nonextensive statistics of entropic parameter $q$.
With the derived expression of the normalized $q$-expectation value, 
we calculated the momentum distribution and the correlation to the order $1-q$ 
as functions of the inverse physical temperature for a free scalar field. 
To the order $1-q$, 
the momentum distribution derived by using the density operator coincides with 
the momentum distribution derived from the entropic measure described with the distribution,  
when the physical temperature equals the temperature in the distribution derived from the entropic measure. 
The correlation depends on the momentums for $q \neq 1$.
The factor two appears in the correlation for the same momentums, 
and indicates that the effects of boson at $q \neq 1$ and those at $q=1$ are similar for the correlation. 
}

\maketitle
\section{Introduction}
The Tsallis nonextensive statistics has been widely applied to various phenomena showing power-like behavior \cite{Book:Tsallis}.
The phenomena with power-like distribution at high energies have been studied, 
such as  momentum distribution \cite{Alberico2009, Cleymans2012, Marques2015, GS2015, Azmi2015, Zheng2016, Lao2016, Cleymans2017-WoC, Cleymans2017, Osada-Ishihara-2017, Bhattacharyya2017-prepri, Yin2017}, 
correlation and fluctuation \cite{Osada-Ishihara-2017, Alberico2000, Ishihara2017-corr-diffmass, Ishihara2017-fluctuation}, 
phase transition \cite{Rozynek2009, Rozynek2016, Ishihara2015, Ishihara2016, Shen2017}, etc.
The statistics has two parameters: one is the temperature $T$ and the other is the entropic parameter $q$. 
The deviation from the Boltzmann-Gibbs statistics is measured by the quantity $1-q$.  
It was shown that the value of $|1-q|$ for the distribution is small at high energies.

The calculation of the expectation value is required to estimate a physical quantity.
Some definitions of the expectation value \cite{Tsallis1998} have been used in the calculation of the quantity in the Tsallis nonextensive statistics. 
The normalized $q$-expectation value is physically relevant, because this expectation value satisfies $\qexpect{1}=1$.  
The normalized $q$-expectation value can be calculated approximately when the measure $|1-q|$ is sufficiently small \cite{Kohyama-Tsallis06}. 
The expectation can be calculated by using $1-q$ expansion for small $|1-q|$ to estimate the physical quantity. \

The momentum distribution was derived by the maximization of the entropic measure described with the distribution 
under constraints \cite{Cleymans2012} which reflect the $q$-expectation value.
This distribution has the temperature $\Tdist$ and the entropic parameter $q$. 
The obtained distribution has been widely used to fit the data, 
and the distribution of $q>1$ derived by the maximization describes well the momentum distributions obtained in the experiments.

It is not always easy to calculate the momentum distribution based on the density operator even for a free field. 
The distribution may be calculated by the $1-q$ expansion when $|1-q|$ is sufficiently small.
The meaning of the parameter $\Tdist$ may be clarified by comparing the distribution calculated by using the density operator 
with the distribution calculated by the maximization of the entropic measure. 
It may be possible to connect the temperature $\Tdist$ and the physical temperature $\Tph$ \cite{Aragao-PhysicaA2003, Eicke-prepri, Kalyana2000}, 
which is related to the variation of the Tsallis entropy and the variation of the internal energies.

The correlation \cite{Andreev1996, Asakawa1999} is widely used to study the phenomena, and is also affected by the distribution. 
Therefore, the correlation should be estimated to clarify the effects of the statistics.
We can calculate directly the quantity with the density operator,  assuming the $q$-expectation value. 
The deviation from the correlation at $q = 1$ is a candidate to study power-like phenomenon.

In this paper, we attempt to derive the expression of the normalized $q$-expectation value of a physical quantity
to the order $1-q$ with the physical temperature.   
The expression is used to find the momentum distribution and the correlation 
for a free scalar field in the Tsallis nonextensive statistics. 
We compare the momentum distribution derived by using the density operator 
with the momentum distribution derived from the Tsallis entropy described by the distribution for small $|1-q|$.

We show the results briefly.  
We derived the normalized $q$-expectation value of a physical quantity to the order $1-q$ 
with the physical temperature in the Tsallis nonextensive statistics.  
With the expression, 
we obtain the momentum distribution and the correlation to the order $1-q$ for a free scalar field.
The equivalence of two momentum distributions to the order $1-q$ is shown 
when the physical temperature equals the temperature introduced in the entropic measure described by the distribution. 
The momentum dependence of the correlation at $q \neq 1$ and the effects of boson are shown. 

This  paper is organized as follows.
In section~\ref{sec:q-1expansion}, 
we review briefly the $q$-expectation value with the density operator  in the Tsallis nonextensive statistics. 
We expand the $q$-expectation value with density operator as a series of $1-q$ 
and obtain the expression of the expectation value to the order $1-q$.
In section~\ref{sec:mom-corr}, 
we obtain the momentum distribution and the correlation to the order $1-q$ 
as functions of the inverse physical temperature for a free scalar field.
The derived momentum distribution is compared with the distribution used 
to fit the experimental data at high energies for small $|1-q|$. 
The last section is assigned for discussion and conclusion. 


\section{Normalized $q$-expectation value for small $|1-q|$}
\label{sec:q-1expansion}
\subsection{Normalized $q$-expectation value in the Tsallis nonextensive statistics}
We start with the $q$-expectation value based on the density operator.
The density operator $\rho$ in the Tsallis nonextensive statistics is defined by 
\begin{align}
\rho := \frac{\rho_u}{\trace{\rho_u}} ,
\quad 
\rho_u := \left[ 1-(1-q) \frac{\beta}{c_q} (H - \qexpect{H}) \right]^{1/(1-q)} , 
\end{align}
where $H$ is the Hamiltonian, $\beta$ is the inverse temperature,  
$q$ is the entropic parameter, $c_q$ is a $q$-dependent constant, 
and $\qexpect{H}$ is the normalized $q$-expectation value of the Hamiltonian.
The normalized $q$-expectation value of a physical quantity $O$ is defined by
\begin{align}
\qexpect{O} := \frac{\trace{\rho_u^q O}}{\trace{\rho_u^q}} .
\end{align}
We adopt the normalized $q$-expectation value in the present study because of physical relevance, $\qexpect{1}=1$. 

The following self-consistent equation should be satisfied from the definition of the expectation value:
\begin{align}
\qexpect{H} = \frac{\trace{\rho_u^q H}}{\trace{\rho_u^q}} , 
\label{eqn:self-consistent}
\end{align}
where $\qexpect{H}$ is included in the right-hand side of eq.~\eqref{eqn:self-consistent}.
The constant $c_q$ and the partition function $Z_q$ should also satisfy the following relation:   
\begin{equation}
c_q = \left( Z_q \right)^{1-q} , 
\label{eqn:cqZq}
\end{equation}
where $Z_q$ is defined by
\begin{equation}
Z_q = \trace{\rho_u} .
\end{equation}
The following calculations are based on these  equations.

\subsection{Normalized $q$-expectation value with the physical temperature for small $|1-q|$}
We calculate the $q$-expectation value of a physical quantity $O$ with small $|q-1|$.
For simplicity, we use the variable $\varepsilon = 1-q$.
We expand $\qexpect{H}$ and $c_q$ with respect to $\varepsilon$ as follows:
\begin{subequations}
\begin{align}
& \qexpect{H} = E_0 - \varepsilon E_1 + O(\varepsilon^2) \label{expand:E} ,\\
& c_q = c_0 - \varepsilon c_1 + O(\varepsilon^2) \label{expand:c} ,\\
& Z_q = Z_0 - \varepsilon Z_1 + O(\varepsilon^2)  \label{expand:Z} .
\end{align}
\end{subequations}
We have the following relations from eq.~\eqref{eqn:cqZq}:
\begin{subequations}
\begin{align}
c_0 &= 1,\\
c_1 & = - \ln Z_0 = -\ln \trace{\exp[-\beta(H-E_0)]}  .
\end{align}
\end{subequations}
We note that the coefficient $c_1$ depends on the volume of the system in general.
The quantity $(\rho_u)^q$ is expanded as follows:
\begin{align}
(\rho_u)^q = e^{\beta E_0} e^{-\beta H} \left\{ 1 + \varepsilon \left[ L_0 + L_1 H + L_2 H^2 \right] + O(\varepsilon^2) \right\}
,
\label{eqn:expansion-of-rho-u^q}
\end{align}
where $L_0$, $L_1$, and $L_2$ are defined by 
\begin{subequations}
\begin{align}
L_0 &= - \beta \left[ (1-c_1) E_0 + E_1 + \frac{1}{2} \beta (E_0)^2 \right] ,\\
L_1 &= \beta \left[ (1-c_1) + \beta E_0 \right] ,\\
L_2 &= -\frac{1}{2} \beta^2  .
\end{align}
\end{subequations}

The $q$-expectation value of $O$ is a function of $\beta$, and is expanded with respect to $\varepsilon$ such as 
\begin{equation}
\qexpect{O}^{(\beta)} = A_{q=1}^{(\beta)} + \varepsilon B_{q=1}^{(\beta)} + O(\varepsilon^2),  
\label{eqn:basic-rel}
\end{equation}
where the argument $\beta$ of the functions is attached as the superscript $(\beta) $ to avoid confusions.
The inverse physical temperature $\betaph$ \cite{Aragao-PhysicaA2003, Eicke-prepri, Kalyana2000} is defined by 
\begin{equation}
\betaph := \frac{\beta}{c_q^{(\beta)}}  . 
\end{equation}
The inverse temperature is written by
\begin{equation}
\beta = c_q^{(\beta)}\betaph =  \left(1 - \varepsilon c_1^{(\beta)} + O(\varepsilon^2) \right)\betaph .
\end{equation}
We obtain the $\varepsilon$ expansion of $\beta$ by using the above equation recursively: 
\begin{equation}
\beta = \betaph  - \varepsilon c_1^{(\betaph)} \betaph+ O(\varepsilon^2) .
\label{eqn:beta-betaph}
\end{equation}
Inserting Eq.~\eqref{eqn:beta-betaph} into Eq.~\eqref{eqn:basic-rel}, we obtain
\begin{align}
\qexpect{O}^{(\beta)}
&= A_{q=1}^{(\betaph)}
+ \varepsilon \left[ B_{q=1}^{(\betaph)} - \left( \frac{\partial  A_{q=1}^{(\betaph)}}{\partial \betaph} \right) c_1^{(\betaph)} \betaph \right]
\nonumber \\& \qquad
+ O(\varepsilon^2) 
.
\label{Obeta-rewritten}
\end{align}
The expansion of $\qexpect{O}^{(\beta)}$ as a function of $\betaph$ to the $O(\varepsilon)$ is obtained 
by using Eqs.~\eqref{eqn:expansion-of-rho-u^q} and \eqref{Obeta-rewritten}:
\begin{subequations}
\begin{align}
\qexpect{O}^{(\beta)}
& = \qoneexpect{O}^{(\beta)} + \varepsilon 
\Big\{  L_1^{(\beta)} \left[\qoneexpect{HO}^{(\beta)} - \qoneexpect{H}^{(\beta)} \qoneexpect{O}^{(\beta)} \right] 
\nonumber \\ & \qquad 
+ L_2^{(\beta)} \left[\qoneexpect{H^2O}^{(\beta)} - \qoneexpect{H^2}^{(\beta)} \qoneexpect{O}^{(\beta)} \right] \Big\} 
\nonumber \\ & \qquad 
+ O(\varepsilon^2) 
\label{final-eq-O-beta} \\
& = \qoneexpect{O}^{(\betaph)} 
+ \varepsilon 
\Big\{ \betaph\left( 1 + \betaph E_0^{(\betaph)} \right) 
\nonumber \\ & \qquad 
\times \left[\qoneexpect{HO}^{(\betaph)} - \qoneexpect{H}^{(\betaph)} \qoneexpect{O}^{(\betaph)} \right] 
\nonumber \\ & \qquad 
- \frac{1}{2} \left( \betaph \right)^2 \left[\qoneexpect{H^2O}^{(\betaph)} - \qoneexpect{H^2}^{(\betaph)} \qoneexpect{O}^{(\betaph)} \right] \Big\} 
\nonumber \\ & \qquad 
+ O(\varepsilon^2) . 
\label{final-eq-O-betaphys}
\end{align}
\end{subequations}
The normalized $q$-expectation value of the physical quantity $O$ is represented 
with the inverse physical temperature $\betaph$ in Eq.~\eqref{final-eq-O-betaphys}.
We note that the $O(\varepsilon)$ term of Eq.~\eqref{final-eq-O-betaphys} does not include $c_1$, 
while the $O(\varepsilon)$ term of Eq.~\eqref{final-eq-O-beta} includes $c_1$ explicitly.

\section{Momentum distribution and correlation for a free scalar field}
\label{sec:mom-corr}
\subsection{Momentum distribution}
We first calculate the number of particles with momentum $\vec{k}$ for a free scalar field.
The momentum distribution based on the density operator for a free scalar field can be obtained with Eq.~\eqref{final-eq-O-betaphys}, 
and the distribution is represented with the inverse physical temperature. 
The Hamiltonian of a free scalar field $H^f$ is given by
\begin{align}
H^f = \sum_{\vec{l}}  \omega_{\vec{l}}  a^{\dag}_{\vec{l}}  a_{\vec{l}}  ,
\end{align}
where $\omega_{\vec{l}}$ is the energy of a particle with momentum $\vec{l}$. 
The operator $a_{\vec{l}}$ is  the annihilation operator which satisfies 
\begin{align}
\left[ a_{\vec{k}}, a^{\dag}_{\vec{l}} \right] = a_{\vec{k}}a^{\dag}_{\vec{l}}  - a^{\dag}_{\vec{l}}  a_{\vec{k}} = \delta_{\vec{k},\vec{l}} . 
\end{align}

We obtain the momentum distribution with vanishing chemical potential from Eq.~\eqref{final-eq-O-betaphys}. 
The number of the particles with momentum $\vec{k}$ is given by
\begin{align}
N_{q; \vec{k}} 
&= \qexpect{a_{\vec{k}}^{\dag} a_{\vec{k}}}^{(\beta)}
\nonumber \\  
&= \qoneexpect{a_{\vec{k}}^{\dag} a_{\vec{k}}}^{(\betaph)} 
  + \varepsilon \Big\{ \betaph\left( 1 + \betaph E_0^{(\betaph)} \right) 
\nonumber \\  & \quad \times 
  \left[\qoneexpect{H^f a_{\vec{k}}^{\dag} a_{\vec{k}}}^{(\betaph)}  - \qoneexpect{H^f}^{(\betaph)} \qoneexpect{a_{\vec{k}}^{\dag} a_{\vec{k}}}^{(\betaph)} \right] 
\nonumber \\ & \quad 
- \frac{1}{2} \left( \betaph \right)^2 
\left[\qoneexpect{\left(H^f\right)^2 a_{\vec{k}}^{\dag} a_{\vec{k}}}^{(\betaph)} 
  \nonumber \right. \\  & \qquad \left.
   - \qoneexpect{\left(H^f\right)^2}^{(\betaph)} \qoneexpect{a_{\vec{k}}^{\dag} a_{\vec{k}}}^{(\betaph)} \right] 
  \Big\} 
+ O(\varepsilon^2) , 
\label{eqn:Nk}
\end{align}
where we define the functions $\Lambda_n^{(\beta)}$ and $\Lambda_{n,\vec{k}}^{(\beta)}$ \cite{Ishihara:preprint}:
\begin{subequations}
\begin{align}
& \Lambda_{n}^{(\beta)} := \trace{ \exp(-\beta H^f) \left(H^f\right)^n } ,
\\
& \Lambda_{n,\vec{k}}^{(\beta)} := \trace{ \exp(-\beta H^f ) \left( H^f \right)^n a^{\dag}_{\vec{k}} a_{\vec{k}}} .
\end{align}
\label{Lambda:main}
\end{subequations}
The normalized $q$-expectation values given in Eq.~\eqref{eqn:Nk} are represented with $\Lambda_n^{(\betaph)}$ and $\Lambda_{n,\vec{k}}^{(\betaph)}$.
The number of the particle with momentum $\vec{k}$ is obtained by using the explicit forms
of $\Lambda_n^{(\beta)}$ and $\Lambda_{n,\vec{k}}^{(\beta)}$ given in the appendix~\ref{app:sec:trace}:
\begin{align}
N_{q;\vec{k}} &= 
\frac{1}{\left( \exp \left(\betaph \omega_{\vec{k}} \right) - 1 \right)}
+ 
\varepsilon \left\{ 
\frac{\left(\betaph \omega_{\vec{k}} \right) \exp \left(\betaph \omega_{\vec{k}} \right) }
{\left( \exp \left(\betaph \omega_{\vec{k}} \right) - 1 \right)^2}
    \right. \nonumber \\ & \left. \qquad 
- 
\frac{\left(\betaph \omega_{\vec{k}} \right)^2 \exp \left(\betaph \omega_{\vec{k}} \right) \left( \exp \left(\betaph \omega_{\vec{k}} \right) + 1 \right)}
{2 \left( \exp \left(\betaph \omega_{\vec{k}} \right) - 1 \right)^3}
 \right\} 
\nonumber \\ & \qquad 
+ O(\varepsilon^2) . 
\label{eqn:number-with-mom-k}
\end{align}

Next, we compare eq.~\eqref{eqn:number-with-mom-k} with the momentum distribution given in ref.~\cite{Cleymans2012}. 
The momentum distribution is obtained from $N_{\vec{k}}$ by taking the large volume limit,  
and the distribution for $\betaph \omega_{\vec{k}} \gg 1$ is 
\begin{align}
\frac{1}{V} \frac{d^3N_q}{d\vec{k}^3} 
&
= \frac{\exp\left( - \betaph \omega_{\vec{k}} \right)}{(2\pi)^3} 
\left\{
 1 + \varepsilon \left[ \left(\betaph \omega_{\vec{k}} \right) - \frac{1}{2} \left(\betaph \omega_{\vec{k}} \right)^2 \right] 
\right\} 
\nonumber \\ & \qquad 
+ O(\varepsilon^2) . 
\label{eqn:momentum-dist-for-density-op-Wien}
\end{align}

The momentum distribution was derived by the maximization of the entropic measure under constraints \cite{Cleymans2012}, 
assuming the Tsallis entropy described with the distribution.
The following Tsallis-type distribution is often used to study momentum distributions at high energies.
We add the superscript 'dist' to $N_q$ in order to avoid confusions.
\begin{equation}
\frac{d^3N_q^{\mathrm{dist}}}{d\vec{k}^3} 
= \frac{gV}{(2\pi)^3} \left[ 1 - \varepsilon \left( \betadist (\omega_{\vec{k}} - \mu) \right) \right]^{\frac{1-\varepsilon}{\varepsilon}}
, 
\label{eqn:cleymans-dist}
\end{equation}
where $g$ is the degeneracy factor and $\mu$ is the chemical potential.
The expansion of eq.~\eqref{eqn:cleymans-dist} at $\mu=0$ and $g=1$ to the order $\varepsilon$ is given by 
\begin{align}
&
\left. \frac{1}{V} \frac{d^3N_q^{\mathrm{dist}}}{d\vec{k}^3} \right|_{\mu=0, g=1} 
\nonumber \\ & 
= \frac{\exp \left( - \betadist \omega_{\vec{k}} \right) }{(2\pi)^3} 
\left\{ 1 + \varepsilon \left[ \left( \betadist \omega_{\vec{k}}  \right) - \frac{1}{2} \left( \betadist \omega_{\vec{k}}  \right) ^2 \right] \right\}
\nonumber \\ & \quad
+ O(\varepsilon^2) . 
\label{eqn:momentum-dist-for-distfunc-Cleymans}
\end{align}

The momentum distribution derived from the normalized $q$-expectation value for the free Hamiltonian to the order $\varepsilon$, 
Eq.~\eqref{eqn:momentum-dist-for-density-op-Wien}, coincides with 
the momentum distribution assuming the Tsallis entropy described with the distribution to the order $\varepsilon$, 
Eq.~\eqref{eqn:momentum-dist-for-distfunc-Cleymans}, when $\betaph$ is equal to $\betadist$.

\subsection{Correlation}
We attempt to calculate the following correlation $C_{q; \vec{k}, \vec{l}}$ which is defined by
\begin{align}
C_{q; \vec{k}, \vec{l}} 
:= \frac{\qexpect{a_{\vec{k}}^{\dag} a_{\vec{l}}^{\dag} a_{\vec{k}} a_{\vec{l}}}^{(\beta)}}{\qexpect{a_{\vec{k}}^{\dag} a_{\vec{k}} }^{(\beta)}\qexpect{a_{\vec{l}}^{\dag} a_{\vec{l}} }^{(\beta)}} 
.
\end{align}
The correlation can be calculated with eq.~\eqref{final-eq-O-betaphys},
as the number of the particles with momentum $\vec{k}$ was calculated. 
The correlation is represented by the functions $\Lambda_n^{(\beta)}$, $\Lambda_{n,\vec{k}}^{(\beta)}$, and $\Lambda_{n,\vec{k},\vec{l}}^{(\beta)}$, 
where the function $\Lambda_{n,\vec{k},\vec{l}}^{(\beta)}$ is defined by 
\begin{align}
\Lambda_{n,\vec{k},\vec{l}}^{(\beta)} := \trace{ \exp(-\beta H^f ) \left( H^f \right)^n a^{\dag}_{\vec{k}} a_{\vec{k}} a^{\dag}_{\vec{l}} a_{\vec{l}}} .
\end{align}
The function $\Lambda_{n,\vec{k},\vec{l}}^{(\beta)}$ is given explicitly in the appendix~\ref{app:sec:trace}.
We finally obtain the correlations, $\left. C_{q;\vec{k}, \vec{l}} \right|_{\vec{k} \neq\vec{l}}$ and $C_{q;\vec{k}, \vec{k}}$, to the order $\varepsilon$:  
\begin{subequations}
\begin{align}
& \left. C_{q;\vec{k}, \vec{l}} \right|_{\vec{k} \neq \vec{l}} 
\nonumber \\ & 
= 
1 - \varepsilon 
\left[ \frac{\left(\betaph \omega_{\vec{k}} \right) \exp\left(\betaph \omega_{\vec{k}} \right)}{\left( \exp\left(\betaph \omega_{\vec{k}} \right) - 1 \right)} \right]
\left[ \frac{\left(\betaph \omega_{\vec{l}} \right) \exp\left(\betaph \omega_{\vec{l}} \right)}{\left( \exp\left(\betaph \omega_{\vec{l}} \right) - 1 \right)}\right]
\nonumber \\ & \quad 
+ O(\varepsilon^2) ,
\\
& \left. C_{q;\vec{k}, \vec{k}} \right. = 
2 - 2 \varepsilon 
\left[ \frac{\left(\betaph \omega_{\vec{k}} \right)^2 \exp\left(2 \betaph \omega_{\vec{k}} \right)}{\left( \exp\left(\betaph \omega_{\vec{k}} \right) - 1 \right)^2} \right]
+ O(\varepsilon^2) .
\end{align}
\label{eqn:correlation:results}
\end{subequations}
Equation~\eqref{eqn:correlation:results} shows that the correlation for $q>1 (\varepsilon < 0)$ is larger than that at $q=1$.

\section{Discussion and Conclusion}
\label{sec:discussion-conclusion}
In this study, 
we derived the expression of the normalized $q$-expectation value based on the density operator to the order $1-q$ 
with the physical temperature in the Tsallis nonextensive statistics of entropic parameter $q$.
We calculated the momentum distribution and the correlation as functions of the inverse physical temperature $\betaph$ 
for a free scalar field, using the expression of the $1-q$ expansion.   
We compared the momentum distribution derived by using the density operator 
with the momentum distribution derived from the Tsallis entropy described by the distribution.

The expression of the $q$-expectation value of a physical quantity as a function of the inverse physical temperature
is useful to estimate the effects of the nonextensivity for small $|1-q|$ in the Tsallis nonextensive statistics with density operator. 
It is significant that 
the $q$-expectation value as a function of the inverse physical temperature to the order $1-q$ does not include $c_1$ 
which depends on the system volume generally. 
The expression implies that the phenomena is described with the physical temperature.

The momentum distribution is an increasing function of $q-1$ for $\betaph\omega_{\vec{k}} \gg 1$,
where $\omega_{\vec{k}}$ is the energy of a particle with momentum $\vec{k}$. 
The tail of the momentum distribution, which implies $\betaph\omega_{\vec{k}} \gg 1$, increases as $q$ increases.
For $\betaph\omega_{\vec{k}} \gg 1$, 
the momentum distribution derived from the normalized $q$-expectation value with the density operator for a free scalar field
is equivalent to the momentum distribution derived from the Tsallis entropy described with the distribution, up to the order $1-q$, 
when the physical temperature in the case of the $q$-expectation value with density operator is equal to the temperature as a parameter of the distribution. 
This fact implies that 
the temperature obtained by the fit to the experimental data is the physical temperature.

The correlation for a free scalar field, $C_q(\vec{k}, \vec{l})$, is an increasing function of $q$ around $q=1$.
The correlation for $q \neq 1$ depends on the momentums, while the correlation for $q=1$ is independent of the momentums.
Evidently, 
the correlation is proportional to $(\betaph \omega_{\vec{k}})(\betaph \omega_{\vec{l}})$
when $|1-q|$ is small enough and $\betaph \omega_{\vec{k}}$ is large enough. 
As for the property of boson, the value of the correlation $C_q(\vec{k}, \vec{k})$ is twice the value of $\lim_{\vec{l} \rightarrow \vec{k}} C_q(\vec{k}, \vec{l})$. 
That is, the effects of boson at $q \neq 1$ and those at $q=1$ are similar for the correlation. 

In summary, 
we derived the normalized $q$-expectation value of a physical quantity to the order $1-q$ 
with the physical temperature in the Tsallis nonextensive statistics, 
and obtained the momentum distribution and the correlation for a free scalar field.  
To the order $1-q$, 
the momentum distribution calculated by using the density operator coincides with  
the momentum distribution derived from the entropic measure described with the distribution, 
when the physical temperature equals to the temperature introduced as a parameter in the distribution derived from the entropic measure.
The correlation $C_q(\vec{k}, \vec{l})$ depends on the momentums. 
The factor two appears in $C_q(\vec{k}, \vec{k})$, as the factor appears in the correlation for the Boltzmann-Gibbs statistics.
This factor indicates that the effects of boson at $q \neq 1$ and those at $q = 1$ are similar for the correlation.

We hope that 
this work is helpful to study the physics at high energies 
within the framework of the Tsallis nonextensive statistics with the density operator.

\appendix
\section{Traces}
\label{app:sec:trace}
In this section, we give the expressions of $\Lambda_n^{(\beta)}$, $\Lambda_{n, \vec{k}}^{(\beta)}$, and  $\Lambda_{n, \vec{k}, \vec{l}}^{(\beta)}$ explicitly.
Again, we write the definitions.
\begin{subequations}
\begin{align}
& \Lambda_{n}^{(\beta)} := \trace{ \exp(-\beta H^f) \left(H^f\right)^n } , \\
& \Lambda_{n,\vec{k}}^{(\beta)} := \trace{ \exp(-\beta H^f ) \left( H^f \right)^n a^{\dag}_{\vec{k}} a_{\vec{k}}} , \\
& \Lambda_{n,\vec{k},\vec{l}}^{(\beta)} := \trace{ \exp(-\beta H^f ) \left( H^f \right)^n a^{\dag}_{\vec{k}} a_{\vec{k}} a^{\dag}_{\vec{l}} a_{\vec{l}}} , 
\end{align}
\end{subequations}
where $H^f$ is the Hamiltonian for a free scalar field:
\begin{align}
H^f = \sum_{\vec{l}}  \omega_{\vec{l}}  a^{\dag}_{\vec{l}}  a_{\vec{l}} , 
\qquad \left[ a_{\vec{k}}, a^{\dag}_{\vec{l}} \right] = \delta_{\vec{k},\vec{l}} . 
\end{align}
The functions $\Lambda_{n}^{(\beta)}$ for $n=0,1,2,3$ are given below:
\begin{subequations}
\begin{align}
& \Lambda_0^{(\beta)} =  \prod_{\vec{l}} \left( 1 - e^{- \beta \omega_{\vec{l}}} \right)^{-1}  ,\\
& \Lambda_1^{(\beta)} =  \left[ \sum_{\vec{l}} \frac{\omega_{\vec{l}}}{\left( e^{\beta \omega_{\vec{l}}} -1 \right)} \right] \Lambda_0^{(\beta)}   ,\\
& \Lambda_2^{(\beta)} =  \left\{ 
\left[ \sum_{\vec{l}} \frac{\left( \omega_{\vec{l}} \right)^2  e^{\beta \omega_{\vec{l}}} }{\left( e^{\beta \omega_{\vec{l}}} -1 \right)^2} \right]
+
\left[ \sum_{\vec{l}} \frac{\omega_{\vec{l}}}{\left( e^{\beta \omega_{\vec{l}}}  -1 \right) } \right]^2 
\right\} \Lambda_0^{(\beta)}  ,\\
&  \Lambda_{3}^{(\beta)}  =  
\Bigg\{ 
\left[ \sum_{\vec{l}} \frac{\left( \omega_{\vec{l}} \right)^3 e^{\beta \omega_{\vec{l}}}\left( e^{\beta \omega_{\vec{l}}} + 1 \right) }{\left( e^{\beta \omega_{\vec{l}}} - 1 \right)^3} \right]
\nonumber \\ & \qquad\qquad 
+
3 \left[ \sum_{\vec{l}} \frac{\omega_{\vec{l}}}{\left( e^{\beta \omega_{\vec{l}}} -1 \right)} \right]
\left[ \sum_{\vec{l}} \frac{\left( \omega_{\vec{l}}\right)^2 e^{\beta \omega_{\vec{l}}} }{\left( e^{\beta \omega_{\vec{l}}} - 1 \right)^2}  \right] 
\nonumber \\ & \qquad\qquad  
+ 
 \left[ \sum_{\vec{l}} \frac{\omega_{\vec{l}}}{\left( e^{\beta \omega_{\vec{l}}} -1 \right)} \right]^3
\Bigg\}
\Lambda_{0}^{(\beta)}
. 
\end{align}
\end{subequations}
The functions $\Lambda_{n,\vec{k}}^{(\beta)}$ for $n=0,1,2$ are given below:
\begin{subequations}
\begin{align}
& \Lambda_{0,\vec{k}}^{(\beta)}  =   \frac{1}{\left( e^{\beta \omega_{\vec{k}}} - 1 \right)} \Lambda_0^{(\beta)}  ,\\
& \Lambda_{1,\vec{k}}^{(\beta)}  =   \left\{ 
\frac{\omega_{\vec{k}} e^{\beta \omega_{\vec{k}}}}{\left( e^{\beta \omega_{\vec{k}}} - 1 \right)} 
+ 
 \left[ \sum_{\vec{l}} \frac{\omega_{\vec{l}}}{\left( e^{\beta \omega_{\vec{l}}} -1 \right)} \right]
\right\}
\nonumber \\ & \qquad\qquad \times
\frac{1}{\left( e^{\beta \omega_{\vec{k}}} - 1 \right)} \Lambda_0^{(\beta)}  ,\\
& \Lambda_{2,\vec{k}}^{(\beta)}  =   
\left\{ 
\frac{\left( \omega_{\vec{k}} \right)^2 e^{\beta \omega_{\vec{k}}}\left( e^{\beta \omega_{\vec{k}}} + 1 \right) }{\left( e^{\beta \omega_{\vec{k}}} - 1 \right)^2}
\right. \nonumber \\ & \qquad\qquad
+
\left[ \frac{2 \omega_{\vec{k}} e^{\beta \omega_{\vec{k}}} }{\left( e^{\beta \omega_{\vec{k}}} - 1 \right)}  \right] 
\left[ \sum_{\vec{l}} \frac{\omega_{\vec{l}}}{\left( e^{\beta \omega_{\vec{l}}} -1 \right)} \right]
\nonumber \\ & \qquad\qquad
+ 
 \left[ \sum_{\vec{l}} \frac{\omega_{\vec{l}}}{\left( e^{\beta \omega_{\vec{l}}} -1 \right)} \right]^2
\nonumber \\ & \qquad\qquad \left.
+ 
 \left[ \sum_{\vec{l}} \frac{\left( \omega_{\vec{l}} \right)^2 e^{\beta \omega_{\vec{l}}} }{\left( e^{\beta \omega_{\vec{l}}} -1 \right)^2} \right]
\right\}
\frac{1}{\left( e^{\beta \omega_{\vec{k}}} - 1 \right)} \Lambda_0^{(\beta)}  
. 
\end{align}
\end{subequations}
The functions $\Lambda_{n,\vec{k},\vec{l}}^{(\beta)}$ for $n=0,1,2$ are given below:
\begin{subequations}
\begin{align}
& \left. \Lambda_{0,\vec{k},\vec{l}}^{(\beta)} \right|_{\vec{k} \neq \vec{l}} = 
\left\{ \frac{1}{(e^{\beta \omega_{\vec{k}}} - 1)} \frac{1}{(e^{\beta \omega_{\vec{l}}} - 1)} \right\} \Lambda_0^{(\beta)} 
, \\
& \left. \Lambda_{0,\vec{k},\vec{k}}^{(\beta)} \right. = 
\left\{ \frac{(e^{\beta \omega_{\vec{k}}} + 1)}{(e^{\beta \omega_{\vec{k}}} - 1)^2}  \right\} \Lambda_0^{(\beta)} 
, \\
& \left. \Lambda_{1,\vec{k},\vec{l}}^{(\beta)} \right|_{\vec{k} \neq \vec{l}} 
\nonumber \\ & 
= 
\Bigg\{
\frac{\omega_{\vec{k}} e^{\beta \omega_{\vec{k}}}}{(e^{\beta \omega_{\vec{k}}} - 1)} + \frac{\omega_{\vec{l}} e^{\beta \omega_{\vec{l}}}}{(e^{\beta \omega_{\vec{l}}} - 1)}
+ \left[ \sum_{\vec{p}} \frac{\omega_{\vec{p}}}{(e^{\beta \omega_{\vec{p}}} - 1)} \right] 
\Bigg\}
\nonumber \\ & \quad \times
\frac{1}{(e^{\beta \omega_{\vec{k}}} - 1)} \frac{1}{(e^{\beta \omega_{\vec{l}}} - 1)}  
\Lambda_0^{(\beta)} , \\
& \left. \Lambda_{1,\vec{k},\vec{k}}^{(\beta)} \right. 
= \Bigg\{ 
\frac{\omega_{\vec{k}} e^{\beta \omega_{\vec{k}}} (e^{\beta \omega_{\vec{k}}}+3)}{(e^{\beta \omega_{\vec{k}}} - 1)^3} 
\nonumber \\ & \qquad\qquad
+ 
\frac{(e^{\beta \omega_{\vec{k}}}+1)}{(e^{\beta \omega_{\vec{k}}} - 1)^2} \left[ \sum_{\vec{p}} \frac{\omega_{\vec{p}}}{(e^{\beta \omega_{\vec{p}}} - 1)} \right] 
\Bigg\} \Lambda_0^{(\beta)} 
, \\
& \left. \Lambda_{2,\vec{k},\vec{l}}^{(\beta)} \right|_{\vec{k} \neq \vec{l}} 
\nonumber \\ & 
= 
\Bigg\{
\frac{(\omega_{\vec{k}} )^2 e^{\beta \omega_{\vec{k}}}}{(e^{\beta \omega_{\vec{k}}} - 1)^2} 
+ \frac{(\omega_{\vec{l}})^2  e^{\beta \omega_{\vec{l}}}}{(e^{\beta \omega_{\vec{l}}} - 1)^2}
+ \left[ \sum_{\vec{p}} \frac{(\omega_{\vec{p}})^2 e^{\beta \omega_{\vec{p}}}}{(e^{\beta \omega_{\vec{p}}} - 1)^2} \right] 
\nonumber \\ & \quad  
+ \left(
\frac{\omega_{\vec{k}} e^{\beta \omega_{\vec{k}}}}{(e^{\beta \omega_{\vec{k}}} - 1)} + \frac{\omega_{\vec{l}} e^{\beta \omega_{\vec{l}}}}{(e^{\beta \omega_{\vec{l}}} - 1)}
   + \left[ \sum_{\vec{p}} \frac{\omega_{\vec{p}}}{(e^{\beta \omega_{\vec{p}}} - 1)} \right] 
   \right)^2 
\Bigg\} 
\nonumber \\ & \quad  \times 
\frac{1}{(e^{\beta \omega_{\vec{k}}} - 1)} \frac{1}{(e^{\beta \omega_{\vec{l}}} - 1)} 
\Lambda_0^{(\beta)} 
, \\
& \left. \Lambda_{2,\vec{k},\vec{k}}^{(\beta)} \right. 
\nonumber \\ & 
= 
\Bigg\{
\frac{ (\omega_{\vec{k}})^2 e^{\beta \omega_{\vec{k}}}}{(e^{\beta \omega_{\vec{k}}} - 1)^4}   (e^{2\beta \omega_{\vec{k}}} + 8 e^{\beta \omega_{\vec{k}}}  + 3)
\nonumber \\ & \qquad  
+ \frac{ 2  \omega_{\vec{k}}  e^{\beta \omega_{\vec{k}}}  (e^{\beta \omega_{\vec{k}}}  + 3)}{(e^{\beta \omega_{\vec{k}}} - 1)^3}
    \left[ \sum_{\vec{p}} \frac{\omega_{\vec{p}}}{(e^{\beta \omega_{\vec{p}}} - 1)} \right]
\nonumber \\ & \qquad  
+ \frac{(e^{\beta \omega_{\vec{k}}}  + 1)}{(e^{\beta \omega_{\vec{k}}} - 1)^2}
\Bigg(
  \left[ \sum_{\vec{p}} \frac{(\omega_{\vec{p}})^2 e^{\beta \omega_{\vec{p}}}}{(e^{\beta \omega_{\vec{p}}} - 1)^2} \right] 
\nonumber \\ & \qquad\qquad
  + \left[ \sum_{\vec{p}} \frac{\omega_{\vec{p}}}{(e^{\beta \omega_{\vec{p}}} - 1)} \right]^2 
\Bigg)
\Bigg\} \Lambda_0^{(\beta)} 
.
\end{align}
\end{subequations}


\end{document}